\documentclass[12pt]{article}
\pdfoutput=1
\usepackage{graphicx}
\usepackage{amsmath}
\usepackage{amssymb}
\usepackage{units}
\usepackage{heppennames2}

\newcommand\pubnumber{DPF2015-248}
\newcommand\pubdate{\today}
\newcommand\sqrts{\ensuremath{\sqrt{s}}}

\def\pnnl{Fundamental Particle Physics\\
Pacific Northwest National Laboratory, Richland, WA, USA}

\def\Title#1{\begin{center} {\Large #1 } \end{center}}
\def\Author#1{\begin{center}{ \sc #1} \end{center}}
\def\Address#1{\begin{center}{ \it #1} \end{center}}

\newcommand\pubblock{\rightline{\begin{tabular}{l} \pubnumber\\
         \pubdate  \end{tabular}}}
\newenvironment{Abstract}{\begin{quotation}  }{\end{quotation}}
\newenvironment{Presented}{\begin{quotation} \begin{center}
             PRESENTED AT\end{center}\bigskip
      \begin{center}\begin{large}}{\end{large}\end{center} \end{quotation}}
\def\Acknowledgments{\bigskip  \bigskip \begin{center} \begin{large}
             \bf ACKNOWLEDGMENTS \end{large}\end{center}}

\begin{document}
\begin{titlepage}
\pubblock

\vfill
\Title{The Higgs Physics Program at the International Linear Collider}
\vfill
\Author{Jan Strube\\for the ILC detector and physics community}
\Address{\pnnl}
\vfill
\begin{Abstract}
The International Linear Collider (ILC) is a proposed electron -- positron collider with a collision energy of $\sqrts=\unit[500]{GeV}$ in the baseline configuration. The ILC physics program takes full advantage of the fact that the machine can be operated at arbitrary energy from the maximum down to the peak of the $\PZ\PH$ production cross section near $\sqrts=\unit[250]{GeV}$ or below. It will advance our understanding of nature through precision measurements of Standard Model parameters, detailed study of the Higgs sector, and a comprehensive search for new phenomena that extends beyond the purely kinematic reach. This note gives an overview of the ILC Higgs program.
\end{Abstract}
\vfill
\begin{Presented}
DPF 2015\\
The Meeting of the American Physical Society\\
Division of Particles and Fields\\
Ann Arbor, Michigan, August 4--8, 2015\\
\end{Presented}
\vfill
\end{titlepage}
\def\thefootnote{\fnsymbol{footnote}}
\setcounter{footnote}{0}
\section{Introduction}
The discovery of a Higgs boson by the LHC experiments completes the Standard Model of Particle Physics, which has enjoyed tremendous success in collider-based particle physics experiments.
However, it cannot explain a number of established phenomena, like cosmological Dark Matter, the accelerating expansion of the Universe, or the observed Matter -- Anti-Matter asymmetry.
New physical models that can explain at least some of these lead to predictions for the coupling between the Higgs boson and the other known fundamental particles that deviate to varying degrees from Standard Model predictions. Measuring the deviations at the per cent level would show the direction of new physics from the size of the deviations.

\section{The ILC Accelerator}

\begin{figure}
    \centering
    \includegraphics[width=0.5\linewidth]{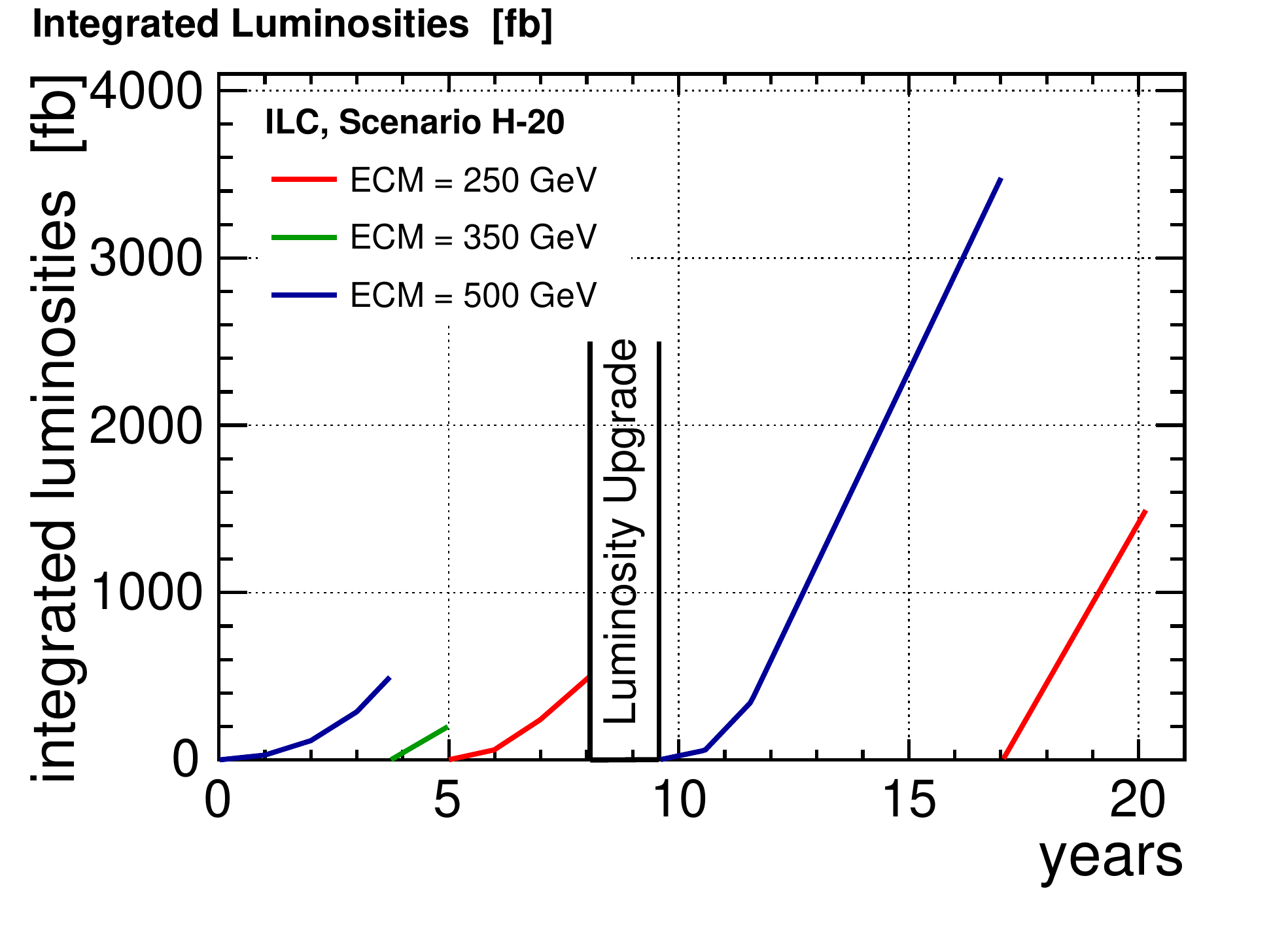}
    \caption{Candidate scenario for operating the ILC in the baseline configuration and the luminosity upgrade.}
    \label{fig:operatingScenario}
\end{figure}
The International Linear Collider is a proposed 31 km electron -- positron accelerator with a center-of-momentum energy of 500 GeV in the baseline.
A candidate site has been identified in the Japanese Kitakami region.
The time lines for different deliverables of the physics program in a number of candidate operating scenarios have been evaluated to guarantee the highest possible impact at the earliest time\cite{Barklow:2015tja}. The preferred candidate for an operating scenario is shown in Figure~\ref{fig:operatingScenario}.

\section{The ILC Detectors}

Detector concepts that can deliver the required precision have been studied in the ILD and SiD groups. They are designed around the particle flow paradigm and feature highly granular calorimeters that are contained in the solenoid. They feature high precision silicon vertex detectors with a trigger-less readout, a low-mass tracking system and nearly $4\pi$ coverage in solid angle. They feature complementary technologies for the main tracker and the hadronic calorimeter. The operating scenario foresees them sharing beam time in a push--pull scenario. The performance of the baseline design has been studied in extensive simulation campaigns. The ability to carry out precision Higgs physics measurements has a large influence on the optimization of the detector design, and studies to improve the understanding of how detector parameters impact the physics performance continue.

\section{The Recoil Technique}
The low background and accurate knowledge of the collision energy at the ILC allows measuring the cross section of $\PZ\PH$ production independently of the $\PH$ decay in a so-called recoil technique. The $\PZ$ decay to a pair of leptons can be reconstructed very cleanly, due to the well-known mass of the $\PZ$ boson and the high momentum resolution of the tracking detectors. A fit to the recoil mass $m_{\text{rec}} = ((\sqrts - E_{\PZ})^2-p_{\PZ}^2)^{1/2}$ allows the measurement of the $\PZ\PH$ cross section without reconstructing decays of the Higgs boson. This technique is applicable at any collision energy, but it has the lowest uncertainty near threshold, where the effect of the finite spread of the collision energy due to beam--beam interaction is smallest.

\section{Higgs Decays to Bottom and Charm Quarks, and to Gluons}
\begin{table}[h]
    \centering
    \begin{tabular}{l|cccc}
        Higgs decay & $\PQb\PAQb$ & $\PQc\PAQc$ & $\Pg\Pg$ & $\PQq\PAQq$ \\
        \hline
        SM prediction & 57.8\% & 2.68\% & 8.56\% & $<0.05\%$ \\
    \end{tabular}
    \caption{Branching ratios of a Standard Model Higgs Boson with a mass of \unit[125]{GeV} to hadrons\cite{Dittmaier:2011ti}. In this table $\PQq$ is an alias for the light quark flavors $\PQs,\PQu,\PQd$. Decays to $\PQt$ quarks are kinematically prohibited.}
    \label{tab:SMHiggsDecays}
\end{table}
The Standard Model prediction for Higgs decays are listed in Table~\ref{tab:SMHiggsDecays}. A large range of models that have survived the discovery and subsequent study of the scalar boson by the LHC experiments predict only per-cent level deviations from these branching ratios.

The Higgs boson decay to a pair of b quark jets has the largest branching ratio, yet is one of the most challenging decays to measure at the LHC due to the large background from b quark jets. The vertex resolution of the detectors is not high enough to tag c quark jets, thus the branching ratio of Higgs decays to c quark jets and gluons cannot be measured at the LHC or its upgrades.

At the ILC, Higgs decays to b and c quark jets, and gluon jets are measured in the signature $\PGn\PGn \PH$ with large missing energy, or $\PZ\PH$ with hadronic or leptonic $\PZ$ decays.
At the ILC, vertices are found from all charged tracks. Charged tracks, reconstructed vertices and neutral calorimeter clusters are combined into two (four) jets in events with large missing energy (hadronic $\PZ$ decays). The jets are then flavor-tagged based on the number of reconstructed vertices and a multi-variate classifier that uses additional kinematic variables. The main background in the channel with missing energy is from hadronic $\PZ$ decays, while the background in the four-jet channel consists mainly of $\PZ\PZ$ and $\PW\PW$ decays. It is mostly reduced by cuts on the invariant mass of jet pairs. Events that remain after selection cuts are fit simultaneously using histogram templates of the flavor tag output.

    The achievable final precision in the ILC physics program is a statistical uncertainty of 0.7\% on the measurement of the b Yukawa coupling, 1.2\% on the measurement of the c Yukawa coupling, and 1.0\% on the measurement to the loop-induced gluon coupling\cite{Asner:2013psa}.

\section{Higgs Decays to $\tau$ Leptons}
The $\tau$ lepton is the heaviest lepton of the Standard Model and as such has the largest branching fraction, with the Standard Model prediction of 6.37\%. This would yield a sizable number of events that can be probed further for CP properties in an angular analysis.
Events are reconstructed in the channels $\PQq\PQq\PGtp\PGtm$ and $\Plp\Plm\PGtp\PGtm$. The analysis includes a $\tau$ jet finder that uses the jet charge to reduce background. Using the  approximation that the visible $\tau$ decay products and the neutrinos are collinear, with no other invisible decay products, the ILC program can achieve a precision on the $\tau$ Yukawa coupling of 1.9\% in the baseline and 0.9\% in a luminosity upgrade (extrapolated from the analysis at \unit[250]{GeV}\cite{Kawada:2014gua}).

\section{Invisible Higgs Decays}
Higgs decays to invisible final states proceed in the Standard Model through the decay $\PH\to\PZ\PZ$, where each $\PZ$ boson in turn decays to a pair of neutrinos. The branching fraction for this decay is 0.1\%. Weakly interacting massive particles, that are of yet unobserved, but are hypothesized as candidates for cosmological dark matter, would lead to a signature of large missing energy in a collider detector. Higgs decays to these particles are kinematically allowed, if their mass is $m \leq m_{\PH}/2$; this would significantly alter the measured branching ratio of invisible Higgs decays.

The recoil mass technique allows the measurement of the branching fraction of these decays in the ILC baseline program with a precision of 0.29\%. This is more than one order of magnitude improvement over LHC predictions and allows for accurate verification of the Standard Model prediction.

\section{Higgs Total Width}
The Standard Model prediction for the natural width of the Higgs boson is \unit[4]{MeV}, too small to be measured directly by analyzing a reconstructed mass distribution, or a production cross section threshold. Under the assumption that the off-shell couplings are identical to the on-shell couplings, the LHC experiments can measure the width of the Higgs boson using off shell decays $\PH\to\PZ\PZ$ with a precision of \unit[22]{MeV}.
At the ILC, the $\PH$ width can be measured in channels where Higgs production and decay are mediated by the same coupling (Equation~\ref{eq:definitionBR}), e.g. a heavy gauge boson coupling. The decay $\PH\to\PW\PW$ can be measured with greater precision than $\PH\to\PZ\PZ$, owing to the larger branching ratio. Using the definition of a branching ratio in Equation~\ref{eq:definitionBR}, one can use a high-precision measurement of a branching ratio, such as $\PH\to\PQb\PQb$ as a crutch to substitute the coupling $g^2_{\PH\PW\PW}$ with the coupling $g^{2}_{\PH\PZ\PZ}$. The latter can be measured in the recoil analysis, independent of the Higgs boson decay.

The $\PZ\PH$ production cross section can be measured with an uncertainty of less than 2\% using the recoil mass technique\cite{Tomita_HadronicHiggs,Yan_leptonicRecoil,Asner:2013psa}.
\begin{equation}
    \label{eq:definitionBR}
    \Gamma_{\PH} = \frac{\Gamma(\PH\to\PW\PW)}{\mathcal{BR}(\PH\to\PW\PW)}\propto\frac{g^2_{\PH\PW\PW}}{\mathcal{BR}(\PH\to\PW\PW)}
\end{equation}
\begin{equation}
     \frac{g^2_{\PH\PW\PW}}{g^2_{\PH\PZ\PZ}}\propto\frac{\sigma_{\PGn\PGn\PH}\times\mathcal{BR}(\PH\to\PQb\PQb)}{\sigma_{\PZ\PH}\times\mathcal{BR}(\PH\to\PQb\PQb)}
\end{equation}

\section{Top Yukawa Coupling}
\begin{figure}
    \centering
    \includegraphics[width=.5\linewidth]{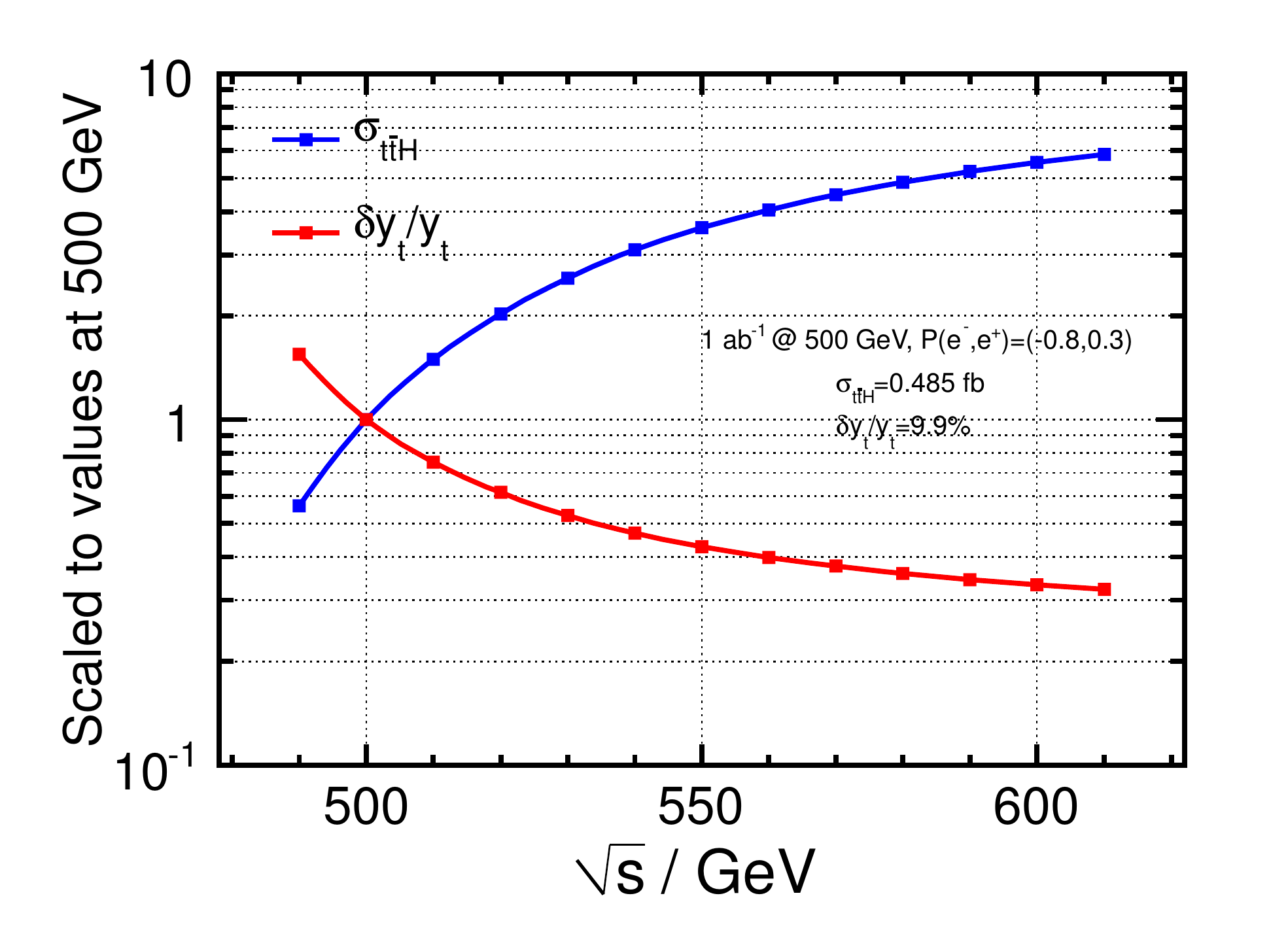}
    \caption{Production cross section of the $\PQt\PAQt\PH$ process and projected measurement uncertainty as function of the collision energy, scaled relative to the value at $\sqrts=\unit[500]{GeV}$.}
    \label{fig:tthError}
\end{figure}
The main challenge for measuring the top Yukawa coupling at the ILC is the low cross section at the collision energy of $\sqrts=\unit[500]{GeV}$. An increase of 10\% in the energy reach of the baseline leads to a nearly four-fold increase of in the $\PQt\PAQt\PH$ production cross section (see Figure~\ref{fig:tthError}).

The analysis strategy of the measurement of the top Yukawa coupling in direct production at the ILC is independent of the collision energy. Isolated tracks from leptonic $\PW$ decays are found and removed from the event. The rest of the event is clustered into four, six or eight jets, depending on whether two, one, or zero isolated leptons were found. Top quarks and the Higgs boson are reconstructed from the jets and leptons using a chi-squared minimization technique with the nominal masses and mass resolution terms found from simulation. Background is reduced using flavor tagging information on the jets, jet size and event shape information. The remaining background is predominantly due to top quark pair production, and to b quark pairs and Z bosons produced in association with top quark pairs.

The projected precision on the top Yukawa coupling in $\unit[1]{ab^{-1}}$ at an ILC upgraded to $\unit[1]{TeV}$ is 4\%\cite{Price:2014oca}. When combined with the analysis at \unit[500]{GeV}\cite{Yonamine:2011jg}, this projects to a precision of 2\% in the full ILC program.
\section{Higgs Self-Coupling}
The self-coupling term of the Higgs potential can be probed in multi-Higgs production at high energy colliders. The quartic coupling involving triple Higgs production is most likely inaccessible for the foreseeable future due to the small cross section of the process. Double Higgs production can be measured at the ILC in the $\PGn\PGn\PH\PH$ channel. The measurement depends on the excellent performance of b-tagging and jet energy resolution to reduce background from $\PZ\PH$ and double $\PZ$ production. With jet clustering in addition to intrinsic detector resolution being a major source of systematic uncertainty, the current estimate for the achievable uncertainty on the tri-linear self-coupling constant in the baseline program is 27\%\cite{Fujii:2015jha}. An energy upgrade to \unit[1]{TeV} will reduce this uncertainty to $\approx 10\%$.

\section{Conclusions}
\begin{figure}
    \centering
    \includegraphics[width=.5\linewidth]{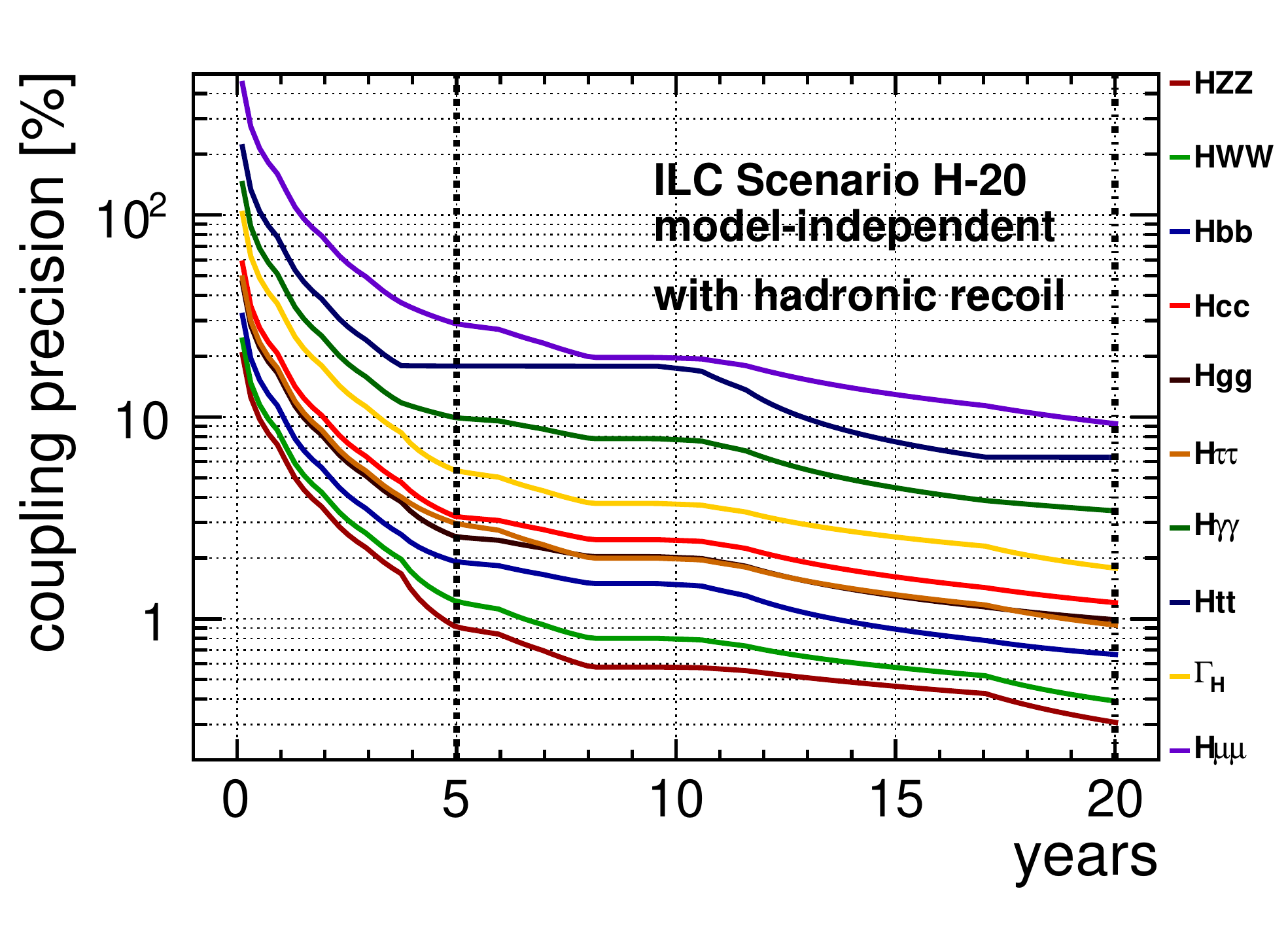}
    \caption{Time development of the precision of the Higgs couplings in scenario ``H'' (Figure~\ref{fig:operatingScenario}).}
    \label{fig:PrecisionDevelopment}
\end{figure}
\begin{table}[t]
\centering
\tiny
\begin{tabular}{l|cccccccc}
     & $\Gamma_{\PH}$ & $g(\PH\PZ\PZ)$ & $g(\PH\PW\PW)$ & $g(\PH\PQb\PAQb)$ & $g(\PH\Pg\Pg)$ & $g(\PH\PQc\PAQc)$,$g(\PH\PQt\PAQt)$ & $g(\PH\gamma\gamma)$ &  $g(\PH\PGt\PGt)$,$g(\PH\PGm\PGm)$  \\
    \hline
    \unit[500]{GeV} ILC (\%)& 0.96 & 0.2 & 0.24 & 0.49 & 0.95 & 1.1 & 3.4 & 0.73 \\
\end{tabular}
\caption{Precision on Higgs width and couplings to SM particles in a global fit to the measurements in the ILC program of the $\sqrts=\unit[500]{GeV}$ baseline configuration (top row) and for the total program including the luminosity upgrade (bottom row). The measurement of $g(\PH\gamma\gamma)$ can be improved to 1.2\% (1.0\%) in the baseline configuration (luminosity upgrade) when taking into account the projected LHC measurements of the ratio of $g(\PH\PZ\PZ)/g(\PH\gamma\gamma)$.}
\label{tab:ilcHiggsGlobalFit}
\end{table}

The ILC baseline program offers a comprehensive picture of the Higgs sector and allows for a self-consistent global fit of all couplings to achieve the greatest precision. Figure~\ref{fig:PrecisionDevelopment} shows the development of the coupling measurements over time in Scenario ``H''. The precision for the measurements of the properties of the LHC Higgs boson in the full ILC program in that scenario is described in the previous sections, with additional improvements of a \unit[1]{TeV} ILC option, where this option leads to a significant improvement. The uncertainties achievable in a the 20-year program of a \unit[500]{GeV} ILC in a global fit using the Snowmass prescription\cite{Dawson:2013bba} are listed in Table~\ref{tab:ilcHiggsGlobalFit}.

This precision is not optional: Different models that can explain some of the obvious discrepancies between the Standard Model and cosmological observations can lead to subtly different predictions in the interactions between the Higgs boson and the known fundamental Standard Model particles. The per-cent-level precision required to distinguish these models can be achieved in the ILC program.
\clearpage
\Acknowledgments
This work was presented on behalf of the global ILC detector and physics community.
\bibliography{ILC_pubs}
\bibliographystyle{plain}
\end{document}